\documentclass[aps,prl,reprint,groupedaddress]{revtex4-1}

\usepackage[T1]{fontenc}
\usepackage[utf8]{inputenc}
\usepackage{graphicx}
\usepackage{amsmath}
\usepackage{amssymb}
\usepackage{color}

\newcommand{\ie}{{\itshape i.\,e.}}

\newcommand{\figref}[1]{Fig.~\ref{#1}}
\newcommand{\beq}{\begin{equation}}
\newcommand{\eeq}{\end{equation}}

\begin{document}

\title{On the Nature of the Debye-Process in Monohydroxy Alcohols: 5-Methyl-2-Hexanol Investigated by Depolarized Light Scattering and Dielectric Spectroscopy}
%\title{The Debye-process in depolarized light scattering of monohydroxy alcohols.}
\author{Jan Gabriel}
\email{Jan.Gabriel@fkp.physik.tu-darmstadt.de}
\author{Florian Pabst}
\author{Andreas Helbling}
\author{Till Böhmer}
\author{Thomas Blochowicz}
\email{thomas.blochowicz@physik.tu-darmstadt.de}

\affiliation{Institut für Festkörperphysik, Technische Universität Darmstadt, 64289 Darmstadt, Germany.}
\date{\today}

\begin{abstract}
The slow Debye-like relaxation in the dielectric spectra of monohydroxy alcohols is a matter of long standing debate. In the present work, we probe reorientational dynamics of 5-methyl-2-hexanol with dielectric spectroscopy and depolarized dynamic light scattering (DDLS) in the supercooled regime. While in a previous study of a primary alcohol no indication of the Debye peak in the DDLS spectra was found, we now for the first time report clear evidence of a Debye contribution in a monoalcohol in DDLS. A quantitative comparison between the dielectric and DDLS manifestation of the Debye peak reveals that while the dielectric Debye process represents fluctuations in the end-to-end vector dipole moment of the transient chains, its occurrence in DDLS shows a more local signature and is related to residual correlations which occur due to a slight anisotropy of the $\alpha$-relaxation caused by the chain formation.
\end{abstract}

\maketitle 

%%%%%%%%%%%%%%%%%%%%%%%%%%%%%%%%%%%%%%
%Introduction
%%%%%%%%%%%%%%%%%%%%%%%%%%%%%%%%%%%%%%

\noindent
The relaxation behavior of monohydroxy alcohols, and in particular the nature of the prominent Debye relaxation peak observed in the dielectric spectra, is a long standing topic \cite{Boehmer:2014}. Initially the Debye model was designed as a model to describe the structural relaxation \cite{Debye1929a}, but later on it turned out that in monohydroxy alcohols the $\alpha$-process is related to a relaxation faster than the Debye peak, and experiments have provided increasing evidence that the Debye contribution is caused by relaxation of transient supra-molecular structures \cite{Boehmer:2014}. In particular the recently proposed model of transient chains \cite{Gainaru2010a}, which form due to H-bonding and, despite their transient nature, effectively lead to a relaxation of an average end-to-end dipole vector, seems promising in that respect.

Although it appeared for a long time that only dielectric experiments show evidence of the Debye process, recently it was identified in the shear mechanical response \cite{Gainaru2014a}
and was also identified in DDLS of several H-bonding liquids \cite{Wang2014a, Hansen:2016}. However, a recent DDLS investigation of
1-propanol with an improved DDLS setup did not show any sign of a Debye contribution, even down to a level of a few percent of the $\alpha$-relaxation amplitude \cite{Gabriel2017a}. This is surprising, first of all because based on the idea of transient chain formation one would expect the observed dynamics to bear some resemblance with the chain  dynamics in unentangled polymer melts \cite{Gainaru2010a}, which in fact can be observed in DDLS \cite{Ding2004a}. Second, even if the picture of polymer-like chain dynamics were not quite adequate,
their formation should at least prevent a fully isotropic $\alpha$-process, and so,
a certain amount of residual correlation should decay on a time scale longer than $\tau_\alpha$. Anticipating that in the latter respect, the situation in primary and secondary alcohols might be different, we present a combined DDLS and BDS study of 5-methyl-2-hexanol (5M2H) in the following, and it turns out that indeed for the secondary alcohol a Debye type process can be identified in light scattering. 

5M2H was purchased from Sigma Aldrich with a purity of 99\,\% and filtered with 20\,nm Watson filters.
The DDLS and BDS experiments were performed with setups already described elsewhere \cite{Blochowicz2013a, Gabriel2015a, Rivera2004a}. Dielectric spectroscopy was performed using a Novocontrol Alpha Analyzer in combination with a time domain dielectric setup. After Fourier transforming the time domain data, combined data sets were obtained covering a frequency range from $10^{-6} - 10^7\,$Hz and a temperature range of $105 - 240\,$K. We note that our dielectric data show good agreement with the data previously published by Bauer et al.~\cite{Bauer2013a}. The photon correlation experiments were performed in vertical-horizontal depolarized geometry. All measurements were taken at a scattering angle of 90 degrees and in a temperature range of $126 - 193\,$K. Particular care was taken to calibrate temperature in the different setups so that an overall accuracy of $\pm 0.5\,$K was achieved.
 
When comparing the correlation functions of DDLS and BDS one has to consider that, besides the fact that both methods probe collective and not microscopic quantities, different correlation functions are involved: DDLS probes the reorientation of the anisotropy tensor of the molecular polarizability \cite{Berne1976a} while BDS probes the collective reorientation of permanent molecular dipole moments, \ie, a vector quantity. Therefore, correlation functions of different Legendre polynomials are probed in both methods: 
\beq
\Phi_\ell(t)=\langle P_\ell(\cos\theta(t))\rangle,
\label{eq:legendre}
\eeq
namely $\ell=1$ for BDS and $\ell=2$ for DDLS,
with $\theta(t)$ being the angle between the positions of the
respective molecular axis at times 0 and $t$. Assuming that the optical anisotropy and the permanent dipole moment are located in the same molecular entity and the considered processes are isotropic to good approximation
a relation between $\Phi_1$ and $\Phi_2$ can be established depending on the geometry of the underlying stochastic process: For example, in case of isotropic rotational diffusion the correlation times are different by a factor of three $\tau^{(1)}/\tau^{(2)} = 3$, in the case of reorientation by random jumps the correlation function becomes independent of $\ell$ \cite{Berne1976a}.

\begin{figure}[t]
\begin{center}\includegraphics[width=8cm]{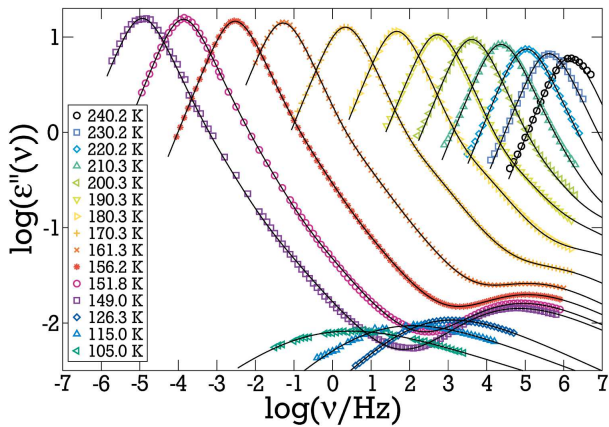}\end{center}
\caption{Frequency dependent dielectric loss data joined with Fourier transformed time-domain 
data of 5-methyl-2-hexanol with fits for Debye, $\alpha$ and $\beta$ relaxation (solid lines).} 
\label{fig:DS-TD-spectren}
\end{figure}
\begin{figure}[t]
%\begin{center}\includegraphics[width=8cm]{./5M2H-g1.eps}\end{center}
\begin{center}\includegraphics[width=8cm]{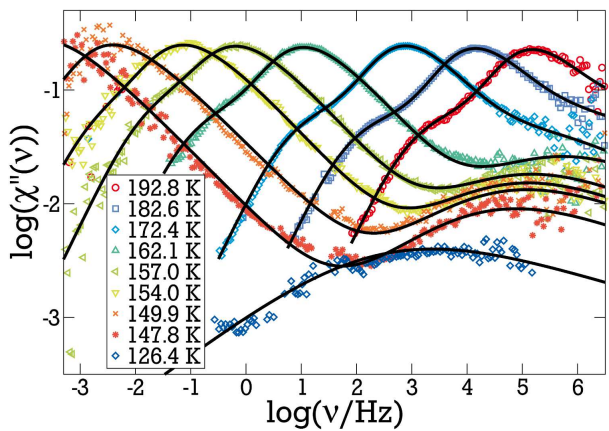}\end{center}
\caption{The Fourier transform of the normalized electric field correlation function $g_1(t)$ from depolarized dynamic light scattering of 5M2H.  Solid lines show the fits for Debye-, $\alpha$- and $\beta$-relaxation as described in the text.
\label{fig:5M2H-g1-pcs-spektren}}
\end{figure}
In general alcohols only show relatively weak anisotropic scattering, so one faces the problem in a photon correlation experiment, that the scattering is partially heterodyne, because small reflections of the incident beam on
the light path, e.g. by sample cell windows or the vacuum shroud, will superimpose as a local oscillator with the weak scattering signal. The treatment of partial heterodyning  and the calculation of the field correlation function $g_1(t)=\left<E_s^*(0)E_s(t)\right>/\left<|E_s|^2\right>$ of the scattered electric field $E_s$ from the intensity correlation function $g_2(t) = \left<I(t)I(0)\right>/\left<I\right>^2$ is described in detail in Refs.~\citenum{Pabst2017a} and \citenum{Gabriel2017a} using the expression:
\begin{equation} 
  g_2(t) = 1+\Lambda C^2|g_1(t)|^2 + 2\Lambda C(1-C)|g_1(t)|,
  \label{eq_Mixed}
\end{equation}
with $\Lambda$ being the independently determined coherence factor and $0 < C \leq 1$ the degree of heterodyning.
To correctly determine $C$, one has to estimate the effect of microcopic dynamics, which causes a decay of $g_1(t)$ on timescales shorter than $t_0 \approx 1\,$ns \cite{Pabst2017a}. For the present data this decay was estimated from Tandem Fabry Perot interferometry measurements at higher temperatures to be $1-g_1(t_0)\approx 0.2$ \cite{Pabst2017unpub}.

Figs.~\ref{fig:DS-TD-spectren} and \ref{fig:5M2H-g1-pcs-spektren}  show dielectric and light scattering results for selected temperatures. In \figref{fig:5M2H-g1-pcs-spektren} the correlation function of the electric field $g_1(t)$ was Fourier transformed to obtain the susceptibility representation. In the light scattering data the contribution of three distinct relaxation processes is directly obvious, with the $\alpha$-relaxation being most prominent in the intermediate time range and a Johari-Goldstein (JG) $\beta$-process at shorter times. At longest times a Debye contribution can clearly be distinguished.  A detailed analysis reveals that basically the same processes are observed by both methods, just the Debye contribution differs significantly in strength, shape and timescale between both techniques. The fit model is the same as the one used earlier for 1-propanol \cite{Gabriel2017a}: $\alpha$- and $\beta$-relaxation are modeled with a distribution of relaxation times, which can be integrated either in frequency or in time domain to yield the dielectric permittivity or the relaxation function, respectively. While this is convenient, we note that the main results are consistently obtained also with more conventional relaxation functions. The expressions for $\alpha$- and $\beta$-relaxation are combined using the Williams-Watts approach \cite{Williams1971a} and finally the Debye-like contribution is added:
\begin{align}
  \Phi(t) & = \Delta\varepsilon_{\! D}\ e^{-(t/\tau_D)^{\beta_{\! D}}} + \nonumber \\ & + \Delta\varepsilon_{\alpha\beta}\ \Phi_\alpha(t)\cdot\left((1-k_\beta) + k_\beta\Phi_\beta(t)\right).
  \label{eq:WW}
\end{align}
We note here, that $\beta_{\! D}$ is included to take account of a possible broadening of the Debye-like process and $k_\beta$ indicates the strength of the JG-relaxation in the Williams-Watts apporach. For more details on the applied relaxation time distributions and the shape parameters see Refs.~\citenum{Gabriel2017a} and \citenum{Blochowicz2003a}. The respective fit results are seen as solid lines in Figs.\ \ref{fig:DS-TD-spectren} and \ref{fig:5M2H-g1-pcs-spektren}. 

\begin{figure}[t]
  \centering
  \includegraphics[width=8cm]{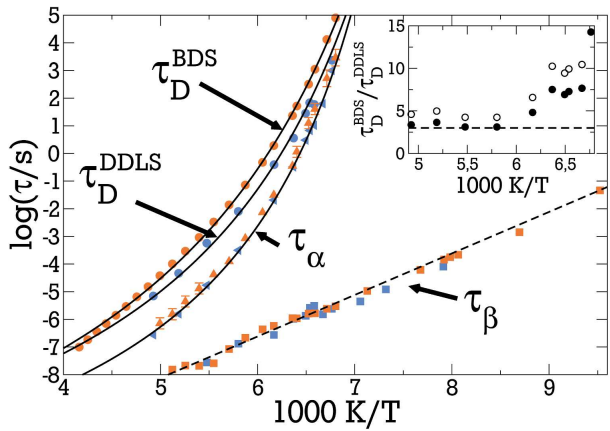}
\caption{Arrhenius plot of relaxation times of 5M2H obtained from BDS (orange) and DDLS data (blue). $\tau_D$ and $\tau_\alpha$ data are interpolated with a Vogel-Fulcher-Tammann law and the $\beta$-process is described by an Arrhenius equation with an activation energy of $E_a/k_B = 3460\,$K.  Unless explicitly indicated, the uncertainties are smaller than the symbol size. The inset shows the ratio of timescales of the Debye contribution for both methods: solid symbols show the ratio $\tau_D^\text{BDS}/\tau_D^\text{DDLS}$, open symbols show the ratio of $\tau_D^\text{BDS}$ and the mean logarithmic correlation time of the DDLS Debye contribution. The dashed line indicates a factor of three as the $\tau^{(1)}/\tau^{(2)}$ ratio in case of rotational diffusion.} 
\label{fig:5M2H-aahrenius-PCS-DS}
\end{figure}
In \figref{fig:5M2H-aahrenius-PCS-DS} we bring together the relaxation times obtained by dielectric and photon correlation spectroscopy. Very similar to the results previously obtained for 1-propanol \cite{Gabriel2017a}, $\alpha$- and $\beta$-relaxation show the same temperature dependent relaxation times in both BDS and DDLS. While $\tau_\alpha(T)$ is described by a Vogel-Fulcher-Tammann law, the JG-relaxation shows an Arrhenius temperature dependence. Within experimental uncertainty the time constants of $\alpha$- and $\beta$-relaxation are identical in both methods. And it turns out that this holds true also for the spectral shape parameters of both relaxations (not shown here), which coincide in the range where a free fit was possible in both data sets. Also, the relative strength of the JG-relaxation does not differ significantly when comparing both methods.

Thus, as the above findings are very similar to what was previously reported in 1-propanol \cite{Gabriel2017a}, a similar conclusion can be drawn about the dielectric and light scattering manifestations of the JG relaxation in 5M2H, namely that if the $\langle P_\ell(\cos\theta(t))\rangle$ correlation functions are directly comparable, which relies on certain assumptions that need careful consideration \cite{Gabriel2017a}, then the JG process in 5M2H cannot be a small angle process,  as often discussed as the motional mechanism behind the JG-relaxation \cite{Vogel2002a}, because in that case its strength would be expected to be different by a factor of three for $\ell=1, 2$ correlation functions \cite{Gabriel2017a}. Instead, reorientation is expected to involve large angle motion, favoring the picture of islands of mobility \cite{Johari1970a} in 5M2H. 

\begin{figure}[t]
  \centering
  \includegraphics[width=8cm]{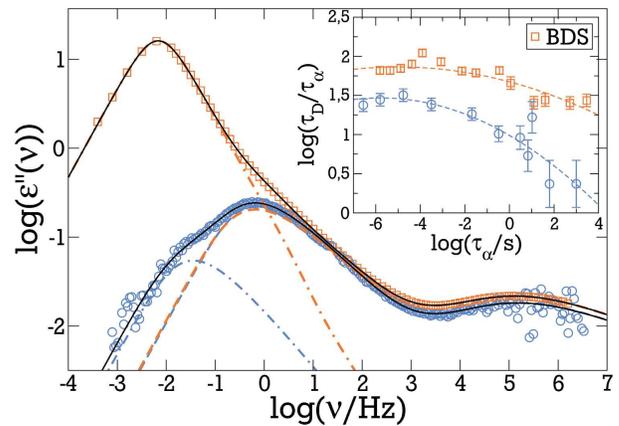}
\caption{Direct comparison of dielectric (orange, squares) and DDLS data (blue, circles) at a similar temperature, demonstrating that $\alpha$- and JG-relaxation appear to be almost identical in both methods with respect to their line shape and time constant (dashed lines), while both methods show a different manifestation of the low-frequency Debye contribution (dash-dotted line).  The inset shows separation between $\alpha$-process and Debye-peak as seen in the ratio $\tau_D/\tau_\alpha$ as a function of the $\alpha$-relaxation time.
\label{fig:5M2H-DS-TD-PCS-vergleich-spektren-157K}}
\end{figure}

While $\alpha$- and JG-relaxation appear almost identical in both methods, the Debye-contribution manifests in a different way, as is demonstrated in \figref{fig:5M2H-DS-TD-PCS-vergleich-spektren-157K}, where the amplitude of the DDLS data is rescaled in order to match the dielectric $\alpha$-/$\beta$-relaxation at a given temperature. It becomes obvious that the Debye contribution is weaker by about two orders of magnitude and faster than its dielectric counterpart. Moreover, it turns out that in DDLS the Debye contribution is broadened: Although the uncertainty in determining the stretching parameter is quite large due to the strong overlap of Debye-like contribution and $\alpha$-process in DDLS the slowest process is characterized by $\beta_D^\text{DDLS}=0.65\pm 0.15$ for all temperatures in \figref{fig:5M2H-g1-pcs-spektren}, while the strong Debye contribution in the dielectric data is well described by $\beta_D^\text{BDS}\approx 1$, in accordance with previous findings for Debye-contributions in BDS and DDLS \cite{Wang2014a, Hansen:2016}. 

One of the hallmarks of the Debye-like process in monohydroxy alcohols is its temperature dependent separation from the $\alpha$-relaxation, which shows a maximum in the intermediate temperature range seen in the ratio $\log(\tau_D/\tau_\alpha)$ around $\tau_\alpha\approx 10^{-4}\,$s \cite{Bauer2013a}. The same ratio is plotted for 5M2H in the inset of \figref{fig:5M2H-DS-TD-PCS-vergleich-spektren-157K}. Qualitatively, it shows the same characteristics in DDLS as in BDS, just that in DDLS the Debye-like contribution is closer to the $\alpha$-relaxation. The ratio of the timescales of the Debye-like contribution in both methods is shown in the inset of \figref{fig:5M2H-aahrenius-PCS-DS}. Solid symbols represent the ratio of $\tau_D^\text{BDS}/\tau_D^\text{DDLS}$, which is around a factor of three and is seen to increase towards lower temperatures. 

Of course a factor of three can be understood in terms of different Legendre polynomial correlation functions Eq.~\eqref{eq:legendre}  $\ell=1, 2$, when considering a process of rotational diffusion, which leads to $\tau^{(1)}/\tau^{(2)} = 3$, and is in accordance with previous findings for the light scattering manifestation of the Debye contribution in other hydrogen bonding systems \cite{Wang2014a,Hansen2016a}. However, as seen in the same inset, the ratio becomes  larger than 3 at lower temperatures. Moreover, when the broadening of the DDLS Debye-contribution is taken into account, an average relaxation time instead of $\tau_D^\text{DDLS}$ has to be considered. Therefore, the ratio with the mean logarithmic relaxation time, calculated as $\ln \langle\tau_D^\text{DDLS}\rangle = (1 - 1/\beta)\cdot \text{Eu} + \ln\tau_D^\text{DDLS}$ \cite{Zorn2002c}, is shown as open symbols in the inset of \figref{fig:5M2H-aahrenius-PCS-DS}. The resulting ratio $\tau_D^\text{BDS}/\langle\tau_D^\text{DDLS}\rangle$ is larger than three at all temperatures. An explanation could be a different effect of cross correlations in both methods, as we point out in the following.

\begin{figure}[t]
  \centering
  \includegraphics[width=8cm]{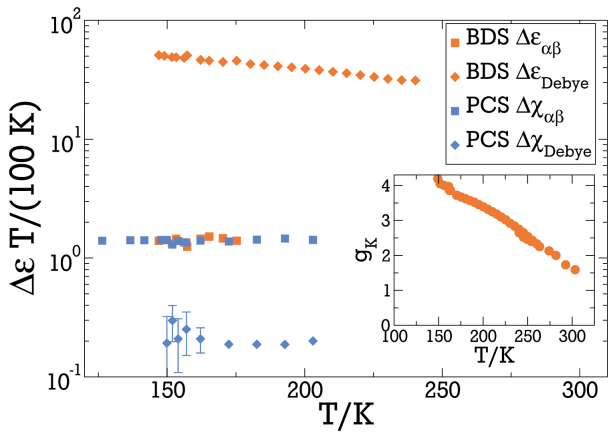}
  \caption{Comparison of the the relative strength of the $\alpha$-/$\beta$-relaxation and the Debye-like contribution in BDS and DDLS. For the dielectric data $T\cdot\Delta\varepsilon_\text{BDS}/100\,$K is plotted to remove the Curie temperature dependence. For the light scattering data $c\cdot\Delta\chi_\text{DDLS}$ is plotted with  $c$ chosen such that $\Delta\chi_{\alpha\beta}$ approximately meets the corresponding BDS data, so that the relative contribution of the Debye-like process becomes directly obvious. The inset shows the Kirkwood correlation factor $g_K$ derived from $\Delta\varepsilon_\text{BDS}$. \label{fig:5M2H-relaxations-staerken-2}}
\end{figure}
\figref{fig:5M2H-relaxations-staerken-2} shows a comparison of Debye- and $\alpha$-/$\beta$-relaxation strength in light scattering and dielectric spectroscopy. It becomes obvious that compared to BDS the Debye-contribution in DDLS is weaker by more than two orders of magnitude. At the same time the remaining temperature dependence in the dielectric $T\cdot \Delta\varepsilon_D$ indicates  that temperature dependent cross-correlations play a role in the dielectric Debye peak. Static cross correlations are usually identified in the Kirkwood correlation factor $g_K$, which is calculated from the Kirkwood-Fröhlich equation \cite{Kirkwood1939a} and is shown in the inset of \figref{fig:5M2H-relaxations-staerken-2}. Like in the case of 1-propanol, $g_K$ is significantly larger than one, indicating that preferentially linear chains are formed by hydrogen bonding, leading to a significant end-to-end-vector dipole moment producing a large dielectric signal. By contrast, for ring-like structures, in which molecular dipole moments would largely cancel each other, $g_K<1$ would be observed \cite{Dannhauser1968a}.

However, intermolecular cross-correlations are also known to influence the dynamics. A theoretical treatment of the problem by Madden and Kivelson \cite{Madden1984a} suggests that this can also be quantified by using the Kirkwood factor of static cross-correlations $g_K$ and a dynamic correlation factor $J_K$ in order to relate a single-molecule correlation time $\tau_s$  with the collective time $\tau_c$:
\begin{equation}
\tau_c \approx \tau_s \cdot g_K/J_k.
\end{equation}
As $J_K\approx 1$ is expected \cite{Madden1984a}, this leads to a slowing down of the collective relaxation if $g_K > 1$. We note that this relation is supported by  experimental findings in hydrogen bonding liquids  as well as by molecular dynamics simulations (see \cite{Weingaertner2004a, MalodeMolina2017a}). In the present context this implies that the relaxation times observed in both methods BDS and DDLS will be affected in a different manner by crosscorrelations quantified by $g_K^\text{BDS}$ and $g_K^\text{DDLS}$, respectively. If one assumes the maximum possible effect of different Legendre polynomials for the single molecule correlation time $\tau_s^{(1)}/\tau_s^{(2)} = 3$, the ratio of characteristic times for the Debye process becomes $\tau_D^\text{BDS} / \tau_D^\text{DDLS} \approx 3\cdot g_K^\text{BDS}/g_K^\text{DDLS}$, where again a possible effect of $J_K$ is neglected. Thus, the ratio of time constants reflects the ratio of cross correlation parameters $g_K^\text{BDS}/g_K^\text{DDLS}$ involved in both methods, which turns out to be significantly larger than 1 and even close to $g_K^\text{BDS}$ for certain temperatures. This indicates, that, while the dielectric Debye contribution is dominated by its collective character the DDLS Debye represents a more local probe of the same dynamics and is closer to the single molecule relaxation time.

This is also reflected in the strength of the Debye contribution in each method: While in BDS it is related to fluctuations of the end-to-end-vector dipole moment, i.e., a collective quantity, where a component of the molecular dipole moment adds up along the contour of the transient chain \cite{Gainaru2010a}, DDLS probes the signature of the Debye contribution in a different manner: Even in case of cross correlations being less important, the formation of transient chains restricts fully isotropic reorientation during $\alpha$-process, leading to some remaining correlation at times $t>\tau_\alpha$. The decay of this remaining level of correlation can then be seen as Debye-like contribution at later times. As such a process is more local in nature than the dielectric Debye it will reflect the heterogeneity of local environments to a larger extent, leading to a significant broadening of the Debye contribution in DDLS, as is observed experimentally. Of course this local decay of remaining correlation is expected to be related to the OH-bond lifetime \cite{Weingaertner2004a}, which is known to typically be on a timescale in between the $\alpha$-relaxation and the dielectric $\tau_D$ \cite{Boehmer:2014}, but will not be identical with it, as cross correlations still do play a role.

The appearance of a Debye-like contribution in DDLS requires, that some correlation is left in an $\ell=2$ correlation function at times longer than $\tau_\alpha$. In primary alcohols like 1-propanol, where so far no Debye contribution could be identified in DDLS on the level of a few percent of the $\alpha$-relaxation amplitude \cite{Gabriel2017a}, the position of the OH group at the end of the alkyl chain may be the reason. Unlike the dipole moment, which is located close to the OH group, the anisotropic polarizability is formed along the alkyl chain, which may have considerable freedom of motion, so that the $\langle P_2(\cos\vartheta)\rangle$ correlation function fully decays via the $\alpha$-process. When this freedom is more restricted due to steric hindrance within a transient chain, as can be envisaged for an OH group position closer to the center of the molecule, some correlation remains leading to an observable Debye-like contribution in DDLS. This conclusion is supported by recent findings in the primary alcohol 2-ethyl-1-hexanol, where again no Debye peak is discernible in DDLS, and in the secondary alcohol 3-methyl-2-butanol, where a pronounced Debye contribution is seen in the data, as shown in the supplemental material [URL here].

In summary we report a clearly discernible Debye-like contribution in the DDLS correlation functions of 5M2H. The strength, timescale and broadening of this contribution indicates, that while the dielectric Debye process reflects fluctuations in the end-to-end vector dipole moment of transient chains, cross correlations are less important for the DDLS Debye contribution, which shows a more local character. How such a scenario is affected in the case when transient chains predominantly form closed-loop structures, and whether or not an underlying single molecule relaxation time is identical with the OH bond lifetime has to be left open for further studies. 

\begin{acknowledgments}
Financial support by the Deutsche Forschungsgemeinschaft under Grant No. BL 923/1 and within FOR 1583 under Grant No. BL 1192/1 is gratefully acknowledged. 
\end{acknowledgments}

%\bibliographystyle{achemso} 
%\bibliography{paper-5M2H,/home/thomas/Dropbox/BibTeX/glasses,/home/thomas/Dropbox/BibTeX/ionicliquids}

%

\end{document}